\documentclass[aps,prb,reprint,groupedaddress,showpacs,showkeys,amsmath]{revtex4-1}
\usepackage{graphicx}
\usepackage{dcolumn}
\usepackage{bm}

\begin{document}

\title{Efficient ``on-the-fly'' calculation of Raman spectra from \textit{ab-initio} molecular dynamics: Application to hydrophobic/hydrophilic solutes in bulk water}

\author{Pouya Partovi-Azar}
\affiliation{Department of Chemistry, University of Paderborn, Warburger Str. 100, D-33098 Paderborn, Germany}

\author{Thomas D. K\"{u}hne}
\affiliation{Department of Chemistry and Institute for Lightweight Design with Hybrid Systems, University of Paderborn, Warburger Str. 100, D-33098 Paderborn, Germany}
\email{tdkuehne@mail.uni-paderborn.de}

\begin{abstract}
We present a computational method to accurately calculate Raman spectra from first principles with an at least one order of magnitude higher efficiency. This scheme  thus allows to routinely calculate finite-temperature Raman spectra ``on-the-fly'' by means of \textit{ab-initio} molecular dynamics simulations. To demonstrate the predictive power of this approach we investigate the effect of hydrophobic and hydrophilic solutes in water solution on the infrared and Raman spectra. 
\end{abstract}

\maketitle

\section{Introduction}

\emph{Ab initio} simulations of vibrational spectra, where the electronic degrees of freedom are explicitly taken into account, often provides important insights into the structure and dynamics of complex systems. Therefore, computer simulations nowadays represent an invaluable tool to rationalize and complement experimental measurements. The key quantity to compute the vibrational spectrum of a system is its dipole moment, which is the tendency of charge distribution towards inhomogeneity. From this point of view, Maximally Localized Wannier Functions (MLWFs) are particularly useful, since they allow to partition the total electronic density into individual fragment contributions. \cite{marzari1997wannier} These Wannier functions are defined as
\begin{equation}\label{wf}
w_n({\bf r}-{\bf R})=\frac{V}{(2\pi)^3} \int_{BZ} d {\bf k} \ e^{-i{\bf k}\cdot{\bf R}} \sum_{m=1}^{J} U_{mn}^{({\bf k})} \psi_{m{\bf k}}({\bf r}),
\end{equation}
where $\psi_{m{\bf k}}({\bf r})$ are Bloch functions, ${\bf R}$ is a Bravais lattice vector, and $V$ is the real-space primitive cell volume, while the integral is computed over the whole Brillouin zone \cite{wannier1937structure}. The unitary $J \times J$ matrix $U_{mn}^{({\bf k})}$ is periodic with respect to the wave-vector ${\bf k}$, while $\psi_{m{\bf k}}({\bf r})$ are the eigenstates of a system as obtained by an electronic structure method, such as density functional theory (DFT) \cite{jones1989dft}. To compute MLWFs, the total spread functional  
\begin{equation}\label{spread}
S = \sum_n S_n  = \sim _n \left( \left< w_n \left| r^2 \right| w_n \right> - \left< w_n \left| {\bf r} \right| w_n \right>^2 \right),
\end{equation}
is minimized by appropriately chosen unitary rotations $U_{mn}^{({\bf k})}$. \cite{marzari1997wannier} Thereof, one can use the expectation value of the periodic position operator $\hat{{\bf r}}$ in the Wannier representation, in order to find the centers of the localized functions for arbitrary symmetries. \cite{marzari1998wannier,silvestrelli1999maximally,berghold2000general} As such, the polarization of the electronic charge in a crystal, which in general has a periodic continuous distribution, can be unambiguously partitioned into localized contributions. \cite{resta1998quantum,resta2007theory} Having the centers and the associated spreads of the MLWFs and considering each MLWF as a charge distribution in space, it is not only possible to calculate the molecular and/or total dipole moments of a system, but also the corresponding time-correlation functions by means of \emph{ab-initio} molecular dynamics (AIMD) simulations.  The temporal Fourier transform of the autocorrelation between the total dipole moments $\bf{M}(t)$ at different times, $t$, is proportional to the infrared (IR) absorptivity, $\alpha(\nu)$, which after employing the harmonic approximation, can be expressed as \cite{ramirez2004quantum,iftimie2005ab,atkins2010atkins}
\begin{equation}
\alpha(\nu) n(\nu) \propto \int_{0}^{\infty} dt \ e^{i 2\pi \nu t} \left< \dot{{\bf M}}(0)\cdot\dot{{\bf M}}(t) \right>_{\rm cl},
\end{equation}
where $\nu$ and $n(\nu)$ are the frequency and the index of refraction, respectively, while $\left< ... \right>_{\rm cl}$ denotes the ensemble-average in classical statistical mechanics. By applying a periodic electric field, \cite{PhysRevLett.89.157602,PhysRevLett.89.117602,PhysRevB.68.085114} usually using the Berry phase approach, \cite{vanderbilt1993polarization,resta1994polarization} it is possible to obtain the polarizability tensor $\hat{A}$ via 
\begin{equation}\label{pol}
A_{ij} = -\frac{\partial M_i ({\bf E})}{\partial E_j}; \ \ i,j=\{x,y,z\},
\end{equation}
where $E_j$ denotes the Cartesian component $j$ of the applied electric field, while $M_i$ is the $i$th component of the total dipole moment. However, calculating the derivatives numerically based on density functional perturbation theory (DFPT) \cite{giannozzi1994vibrational,baroni2001phonons,PhysRevLett.88.176401} or using higher-order finite difference (FD) methods \cite{vasiliev1997ab} is computationally rather expensive. The mean polarizability $\bar{A}=1/3 \ {\rm Tr}[\hat{A}]$ is the quantity that is usually measured in experiment. With $\bar{A}$ known, one can obtain the isotropic Raman spectrum through the calculation of the autocorrelation between the polarizabilities  \cite{atkins2010atkins,ishiyama2011molecular}
\begin{equation}
\sigma(\nu) \propto \nu \ \int_{0}^{\infty} dt \ e^{i 2\pi \nu t} \left< \bar{A}(0)\bar{A}(t) \right>_{\rm cl}.
\end{equation}
In this paper we present a novel computational technique to efficiently calculate Raman spectra, where the polarizability of a system is represented as a sum over Wannier function polarizabilities denoted as a function of Wannier function volumes, and consequently by their spread.

\section{Wannier polarizability method}

Specifically, it has been shown that the molecular polarizability change linearly with the volume of the electronic cloud around a molecule. \cite{murray1993relationships} Due to the fact that the electronic properties play the main role in quantifying the polarizability, it is reasonable to assume that the total isotropic polarizability of the system can be obtained as a sum over the Wannier function polarizabilities. The essentially same idea has been recently used to combine the quantum harmonic oscillator model with Wannier functions to calculate the non-local dynamic electron correlation due to oscillating charges assigned to MLWFs. \cite{:/content/aip/journal/jcp/139/5/10.1063/1.4816964} As such, the polarizability assigned to $i$th MLWF is given by
\begin{equation}\label{gamma}
A_i = \beta S_i^3,
\end{equation}
where $S_i$ is the spread of the $i$th MLWF and $\beta$ is a proportionality constant. Therefore, the mean polarizability of the system can be written as
\begin{equation}\label{alpha}
\bar{A}= \frac{1}{3} {\rm Tr}[\hat{A}] = \frac{1}{3} \sum_{i}^{N_{\rm WF}} A_i = \frac{\beta}{3} \sum_{i}^{N_{\rm WF}} S_i^3.
\end{equation}
At variance to the FD technique, the numerical effort reduces from six to just one single-point calculation plus an additional MLWF computation. This is to say that our novel MLWF-based method, which hereafter we will refer to as Wannier polarizability (WP) method, is at least five times more efficient than the conventional FD approach. The speed-up with respect to DFPT is similar. \cite{putrino2000generalized}

The parameter $\beta$ is determined by minimizing the Mean Absolute Relative Error (MARE) of the mean polarizabilities with respect to reference calculations using the FD approach. All of our DFT calculations were conducted using the CP2K/\textsc{Quickstep} code \cite{vandevondele2005quickstep} in conjunction with a very accurate TZV2PX Gaussian basis set \cite{vandevondele2007gaussian} and the Perdew-Burke-Ernzerhof exchange-correlation functional \cite{perdew1996generalized} plus a damped interaction potential to approximately account for long-range dispersion interactions. \cite{grimme2010consistent}
\begin{figure} [!t]
\centerline{
\includegraphics[width=0.475\textwidth]{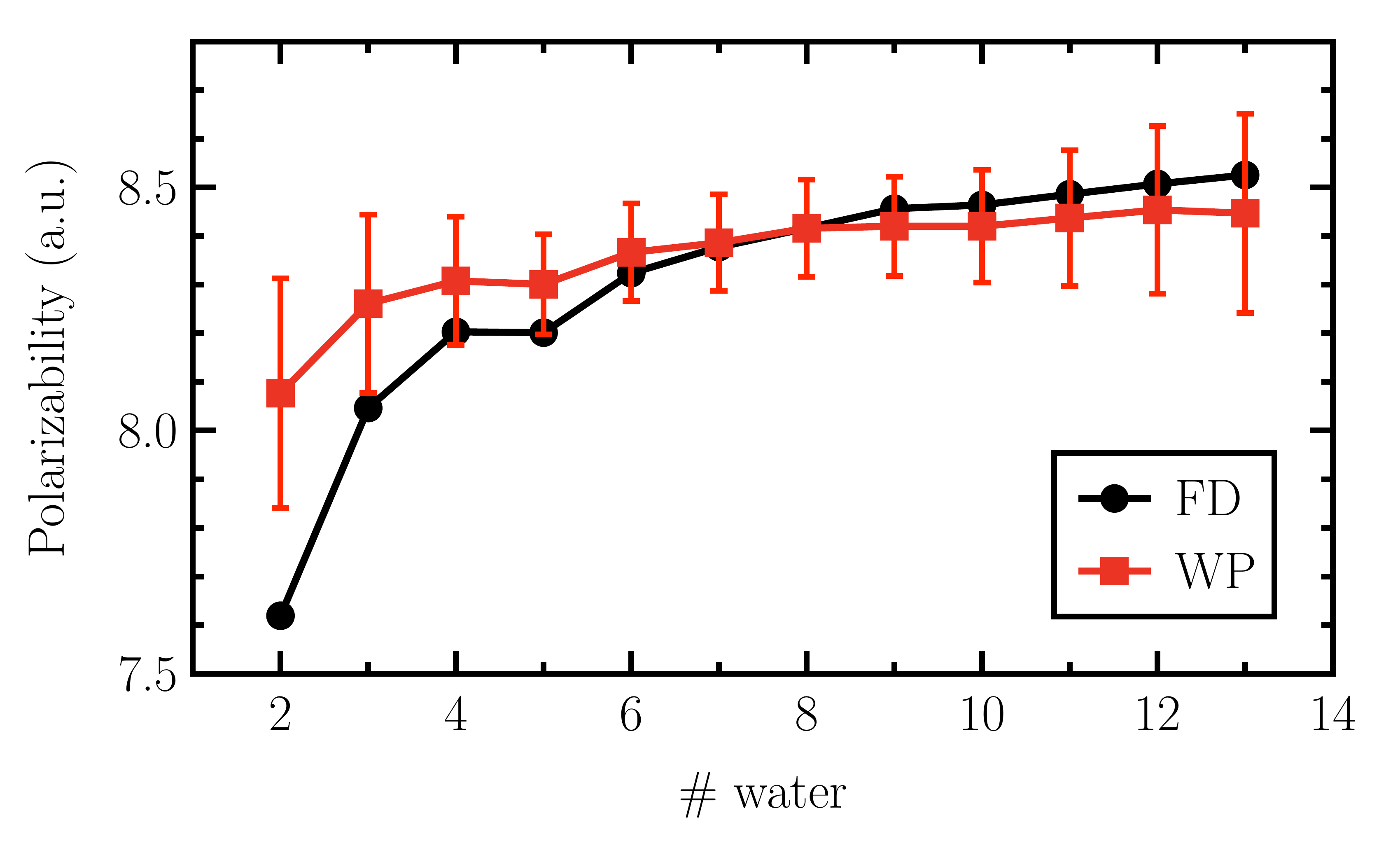}}
\caption{Mean polarizabilities per water molecule for various water clusters containing 2 to 13 molecules as obtained using the FD (black circles) and WP (red squares) methods. The error bars for the red squares are the absolute relative errors at each point.}\label{fig1}
\end{figure}
The mean polarizabilities per molecule for various water clusters containing 2 to 13 molecules as obtained using the FD scheme are shown in Fig.~\ref{fig1}. The structures of the water clusters at Hartree-Fock level of theory were taken from the Cambridge Cluster Database. \cite{doi:10.1021/jp013141b} The obtained results are in very good agreement with previously calculated polarizabilities for the same systems at DFT level of theory. \cite{yang2005dft, ghanty2003polarizability}

The optimized value of $\beta$ for our WP method at minimum MARE is $\beta=0.90$. 
The corresponding isotropic polarizabilities of the water clusters are also displayed in Fig.~\ref{fig1}. Even though the absolute polarizabilities of the FD and WP methods slightly differ for the smallest water clusters, the qualitative behavior with respect to cluster size is similar, which immediately suggest that the deduced Raman spectra most likely differ only in their absolute intensities. 
\begin{figure} [!t]
\centerline{
\includegraphics[width=0.25\textwidth]{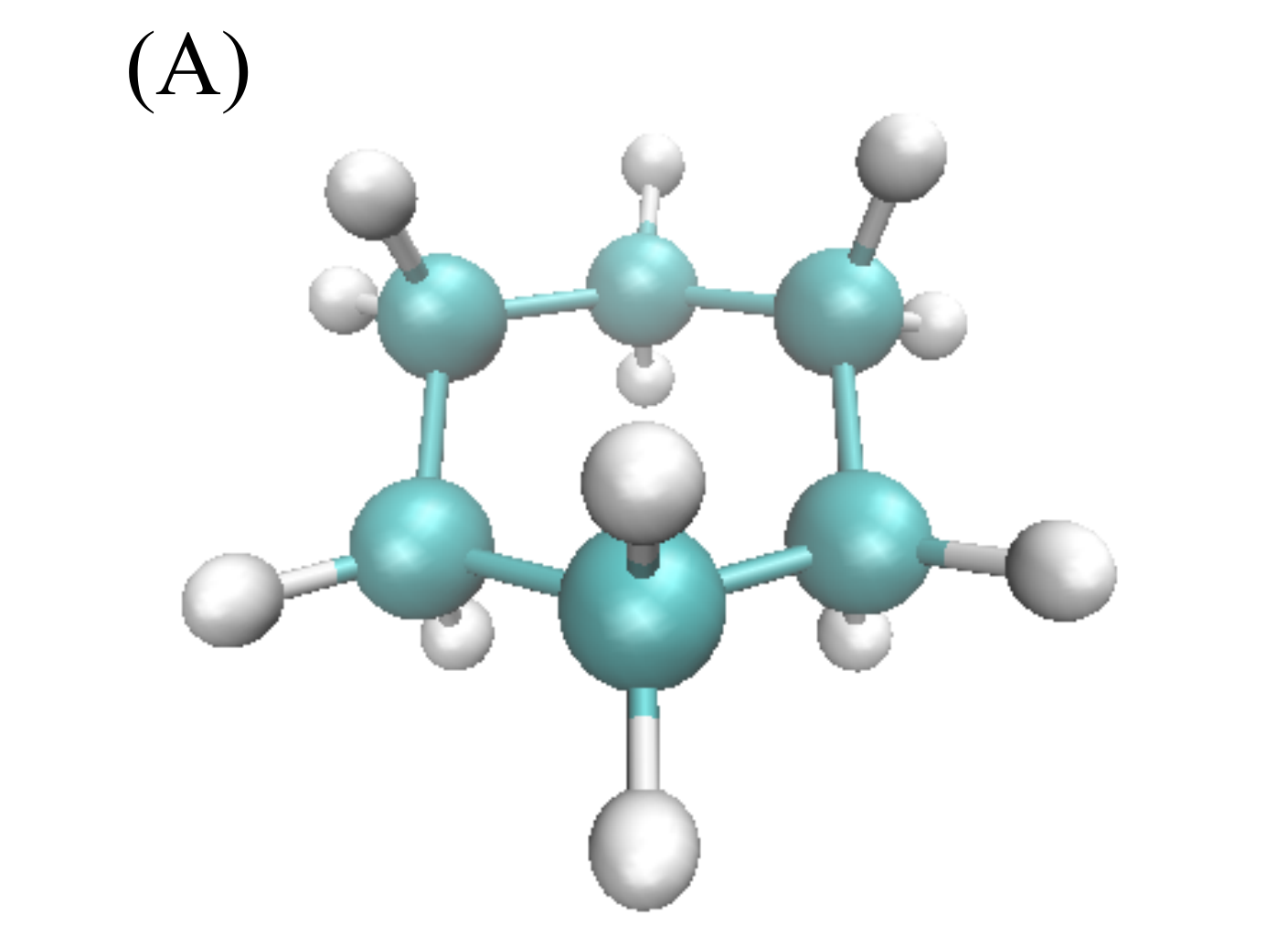}
\includegraphics[width=0.25\textwidth]{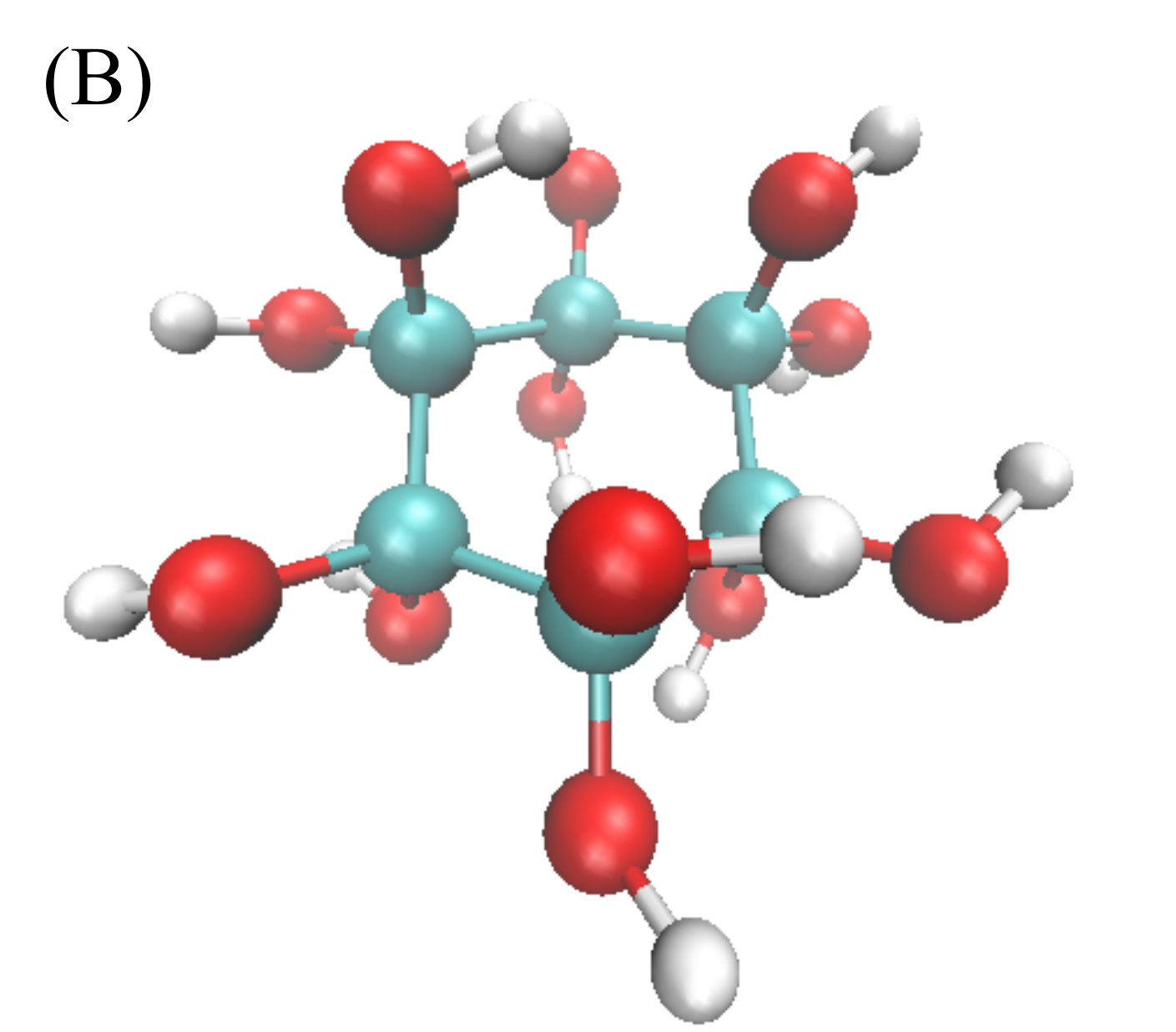}
}
\caption{Atomic configuration of cyclohexane (a) and cyclohexanedodecol (b). Cyan, red and white spheres denote carbon, oxygen and hydrogen atoms, respectively.}\label{fig2}
\end{figure}
To assess the latter, we have compared the Raman spectra of a single cyclohexane molecule in the gas phase, shown in Fig. \ref{fig2}(a), using both FD and WP methods. The spectra were obtained ``on-the-fly'' from simulations using the second-generation Car-Parrinello method, where both the density matrix and $U_{mn}^{({\bf k})}(t)$ were propagated together with the nuclei to further speed-up the calculations. \cite{Kuhne2007, Kuhne2014} However, for each AIMD step, the Wannier functions were re-localized using the scheme of Berghold et al. to obtain genuine MLWFs. \cite{berghold2000general} Nonetheless, together with the WF method, this results in a combined acceleration for the calculation of the Raman spectra of at least an order of magnitude. At first the system was equilibrated in the canonical ensemble at 300~K for 10~ps using a discretized time-step of 0.5~fs, before sampling the polarizabilities in the micro-canonical ensemble for additional 10~ps. In the case of FD approach, at each time-step we applied an external electric field of 0.0001~a.u. intensity along the $\pm x$, $\pm y$ and $\pm z$ directions, to calculate the polarizabilities. 
\begin{figure} [!t]
\centerline{\includegraphics[width=0.475\textwidth]{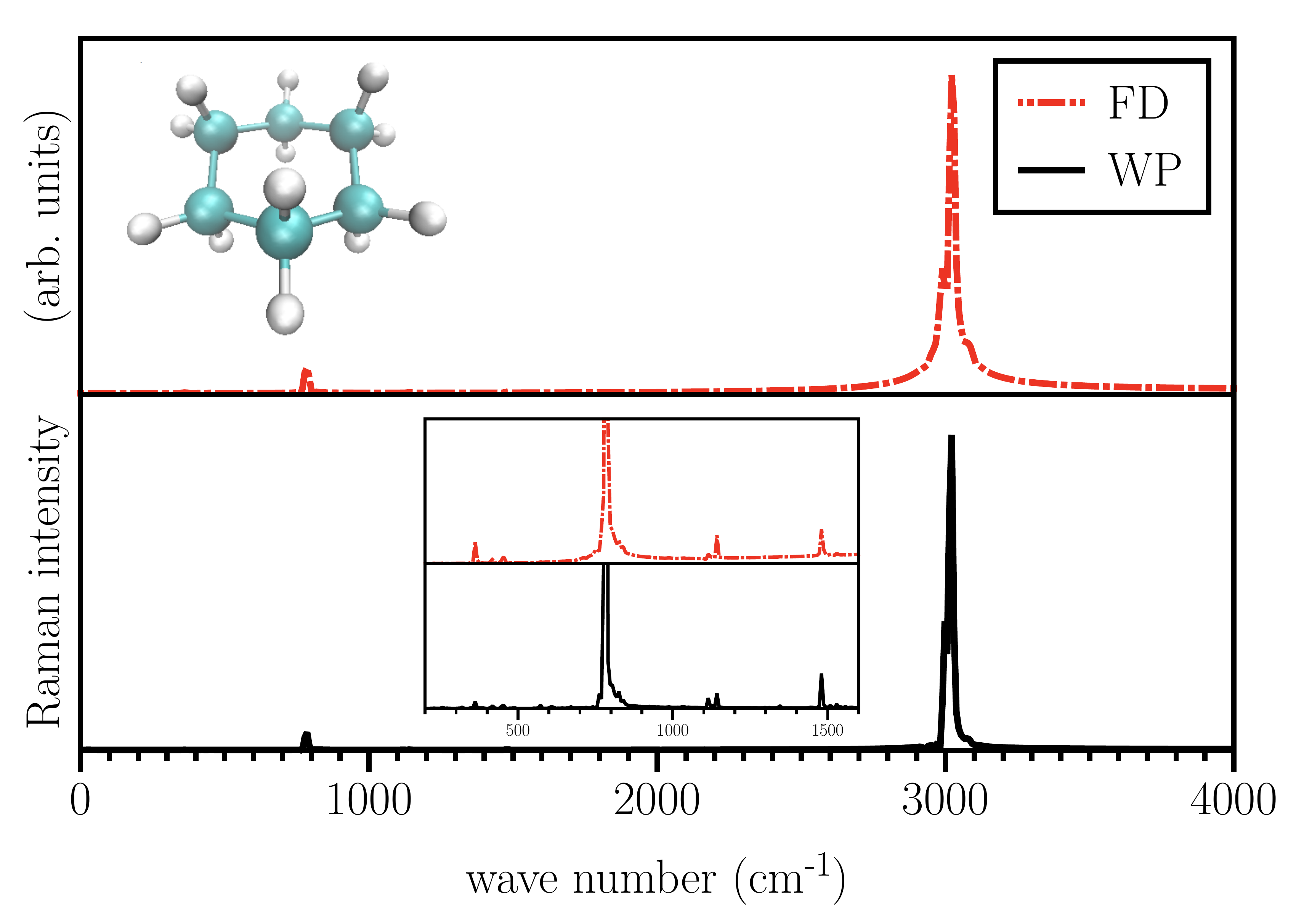}}
\caption{Isotropic Raman spectra of cyclohexane molecule in the gas phase, as obtained by the FD (red dashed line) and WP (black straight line) methods.}\label{fig3}
\end{figure}
The eventual Raman spectra are depicted in Fig.~\ref{fig3}. As can be seen, the agreement between the FD and WP methods is excellent, and no frequency shift can be observed. The peaks around 800 cm$^{-1}$ are typically attributed to C--C stretching and CH$_2$ rocking vibrations, while the peaks around 3000 cm$^{-1}$ are usually assigned to symmetric and asymmetric CH$_2$ stretches. In the inset of Fig.~\ref{fig3} the Raman activity for frequencies less than 1500 cm$^{-1}$ are shown, where the higher bands are due to other CH bending modes, while the lower ones originates from the torsion and deformation of the carbon ring \cite{ito1965raman,wiberg1971vibrational,matrai1985scaled,hirschfeld1986ft,frankland1999molecular,pelletier1999effects,jordanov2003peculiarities,kukura2007femtosecond}. Our simulated spectra are in good agreement with other theoretical and experimental results.

\section{Results and discussion}

\begin{figure} [!t]
\centerline{
\includegraphics[width=0.25\textwidth]{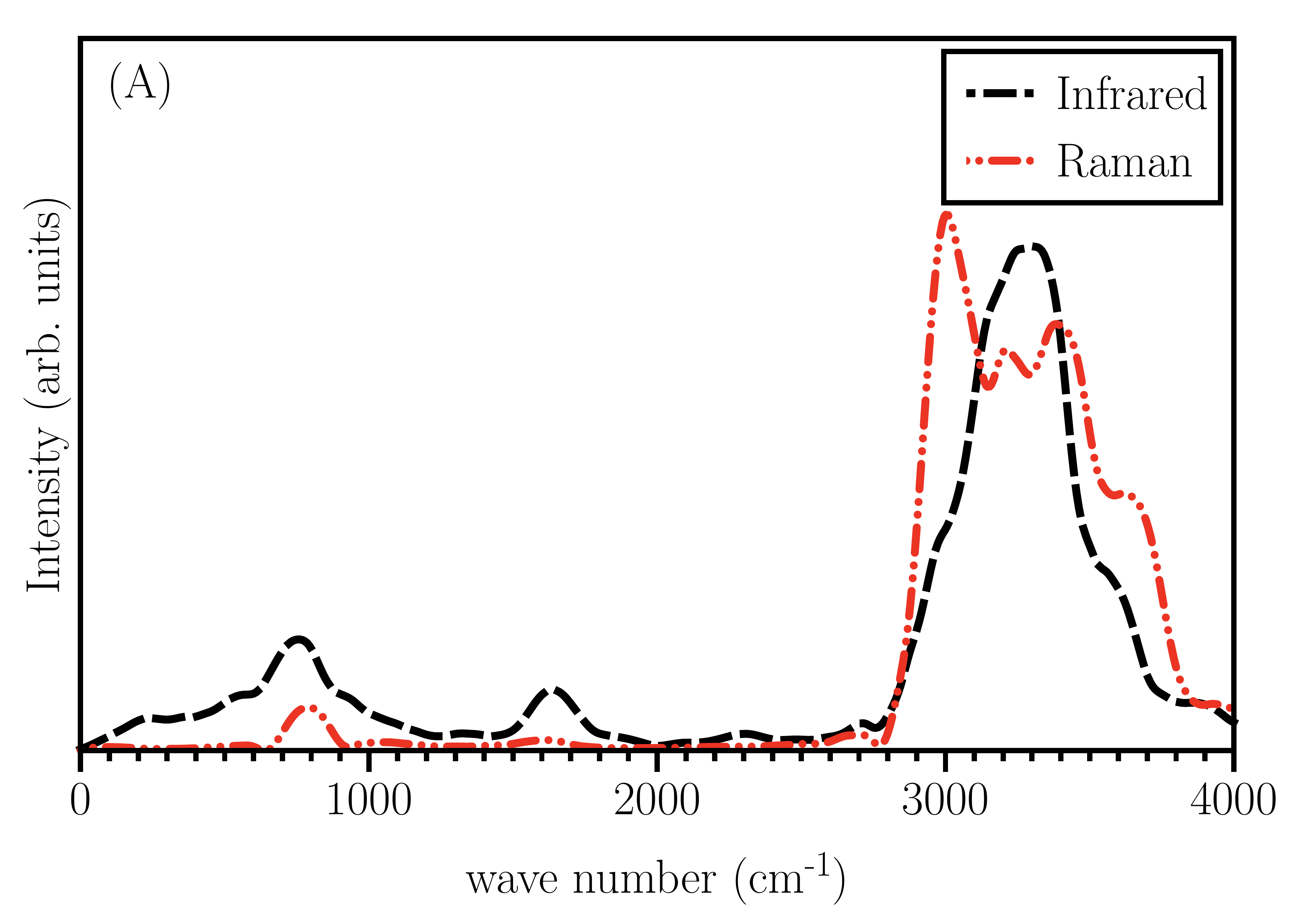}
\includegraphics[width=0.25\textwidth]{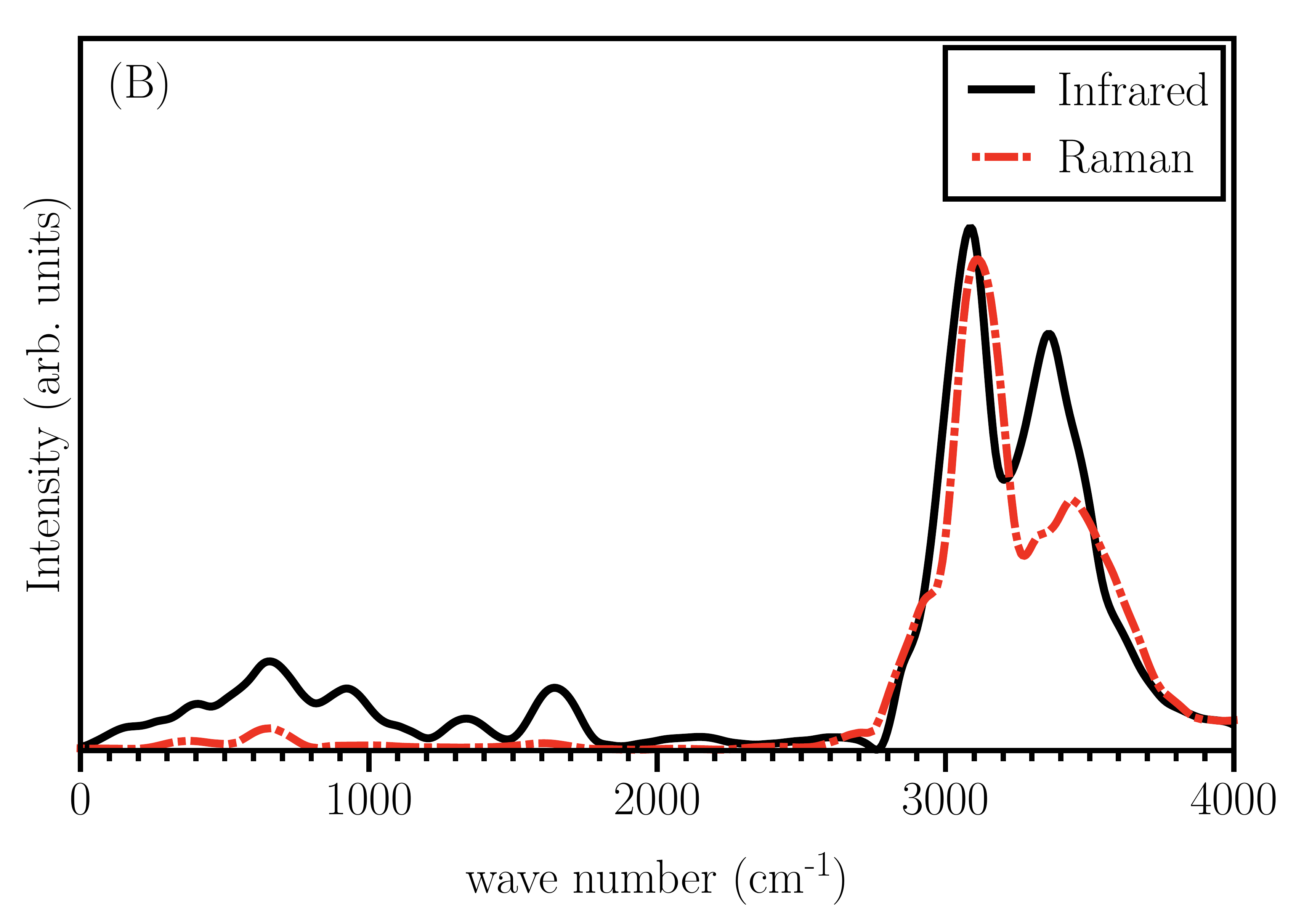}
}
\caption{IR and Raman spectra of CW (a) and COHW (b) systems. Black and red curves show infrared and Raman spectra, respectively.}\label{fig5}
\end{figure}

We demonstrate our novel WP method by investigating the effect of hydrophobic and hydrophilic molecules on the IR and Raman spectra of liquid water. To that extend we consider two systems in water solution: a single cyclohexane molecule (Fig.~\ref{fig2}(a)) to represent a hydrophobic solute, and the relatively similar, but hydrophilic cyclohexanedodecol molecule (Fig.~\ref{fig2}(b)), where all hydrogen atoms are replaced by OH groups. Hereafter we refer to these two systems as CW and COHW, respectively. Both, CW and COHW contains 128 light water molecules per unit cell, where the water density is set to the experimental value at ambient conditions. Contrary to our previous calculations, a smaller DZVP basis set has been employed. Again, both structures were first equilibrated in the canonical ensemble at room temperature for $\sim$15~ps, before the dipole moments and polarizabilities were sampled ``on-the-fly'' for 10~ps in the micro-canonical ensemble. The simulated IR and Raman spectra of CW and COHW systems are shown in Fig.~\ref{fig5}.
The IR activities for the frequencies below 650 cm$^{-1}$, at $\sim$1600 cm$^{-1}$, as well as the weak peak at $\sim$ 2200 cm$^{-1}$ are due to libration and hydrogen bond bending and stretching, O-H bending, and a combination of O-H bending and libration modes of bulk water, respectively. \cite{marechal1991infrared,carey1998measurement,larouche2008isotope,max2009isotope,sun2013local} The peak at $\sim$800 cm$^{-1}$ in the IR and Raman spectra of the CW and COHW systems originate most likely from C-C stretching modes of the carbon ring. \cite{ito1965raman,wiberg1971vibrational,matrai1985scaled,hirschfeld1986ft,frankland1999molecular,pelletier1999effects,jordanov2003peculiarities,kukura2007femtosecond} 
Moreover, IR activities in the frequency range 950-1100 cm$^{-1}$ can be assigned to C-O stretching modes of cyclohexanedodecol, while the peak at $\sim$1350 cm$^{-1}$ seems to be due to its C-O-H bending mode. \cite{antony2005anharmonic}
By comparison of Fig.\ref{fig5}(a) with Fig.\ref{fig5}(b) we attribute the IR-active peak at $\sim$3000 cm$^{-1}$ of the CW system to C-H stretching modes of the cyclohexane molecule. The remaining frequencies above above 3000 cm$^{-1}$ are the O-H stretching modes. \cite{Kuhne2009static, zhang2013vibrational} However, at variance to bulk water, we observe a generally larger splitting between the symmetric and asymmetric stretching modes, which immediately suggests that the asymmetry of the hydrogen-bond network is more pronounced due to the presence of the solute. \cite{kuhne2013electronic, zhang2013vibrational, kuhne2014nature} In the case of the hydrophobic cyclohexane molecule the effect is even more pronounced. Moreover, for the CW system the so-called dangling O-H bond peak at $\sim$3650 cm$^{-1}$ is more distinct and gives rise to a shoulder as can be seen in Fig.~\ref{fig5}(a). \cite{PhysRevLett.70.2313,du1994surface,PhysRevLett.100.096102, nihonyanagi2011unified} We believe that the latter is a consequence of fleetingly broken hydrogen-bonds of the distorted hydrogen-bond network which is spanned around the hydrophobic solute. \cite{chandler2005interfaces}

\section{Conclusions}

In summary, we have presented a novel method that allows to efficiently calculate IR and in particular Raman spectra ``on-the-fly'' within AIMD simulations. To that extend we exploit the fact MLWFs, which are at the core of this new approach, can be utilized to partition the charge distribution of the system into localized fragments. Therefore, the total isotropic polarizability can be calculated as a sum over the Wannier polarizabilities, which are assumed to be proportional to its volume and determined by its spread. Together with an extension of the second-generation Car-Parrinello method to propagate $U_{mn}^{({\bf k})}(t)$ along with the nuclei, followed my a re-localization to obtain genuine MLWFs, a speed-up of one order of magnitude has been observed. Using this approach, we calculated IR and Raman spectra for cyclohexane and cyclohexanedodecol solutes in ambient bulk water. We found that the former hydrophobic solute give rise to a shoulder at around $\sim$3650 cm$^{-1}$, which is due to momentarily dangling O-H bonds. In any case, we conclude by noting that this development facilitates to routinely calculate finite temperature spectra with only minimal extra computational cost.

\bibliography{PW}

\end{document}